\documentstyle[pre,aps,preprint]{revtex}
\tightenlines
\def\e{\epsilon}
\def\s{\sigma}
\def\a{\alpha}
\def\lb{{\langle}}
\def\rb{{\rangle}}

\def\g{\gamma}
\def\Om{\Omega}
\def\om{\omega}
\def\rv{{\bf r}}
\def\vv{{\bf v}}
\def\kv{{\bf k}}
\def\bar{\overline}
\def\half{{1\over 2}}
\def\beq{\begin{equation}}
\def\eeq{\end{equation}}
\begin{document}

\centerline{\bf Molecular Dynamics Study of Long-Lived Structures}
\centerline{\bf in a Fragile Glass Forming Liquid}
\medskip
\centerline{Gregory Johnson,$^1$ Andrew I.\ Mel'cuk,$^2$ Harvey
Gould,$^1$ W.\ Klein,$^3$ and Raymond D.\
Mountain$^4$}
\centerline{$^1$Department of Physics, Clark University,
Worcester, MA 01610}
\centerline{$^2$Polymer Science and
Engineering Department,}
\centerline{University of Massachusetts, Amherst, MA
01002}
\centerline{$^3$Department of Physics, Boston University, Boston,
MA 02215}
\centerline{$^4$Physical and Chemical Properties Division,}
\centerline{National Institute of Standards and Technology,
Gaithersburg, MD 20899}
\bigskip
\centerline{\bf Abstract}

\noindent We present molecular dynamics results for a two component,
two-dimensional Lennard-Jones supercooled liquid near the glass
transition. We find that the supercooled liquid is spatially
heterogeneous and that there are long-lived clusters whose size
distribution satisfies a scaling relation up to a cutoff. The
similarity of several properties of the supercooled liquid to those
of a mean-field glass-forming fluid near the spinodal suggests that
the glass transition in the supercooled liquid is associated with an
underlying thermodynamic instability.

\bigskip \bigskip \noindent {\bf 1. Introduction}\hfill\break
\noindent The characterization of supercooled liquids near the
glass transition is an active area of research~\cite{reviews}.
Outstanding unsolved problems include the possible existence of an
underlying thermodynamic glass transition, the history dependence
of the glass properties, and the mechanisms responsible for the
large increase in relaxation times as the glass transition is
approached. In this article we discuss our molecular dynamics
simulations of a two component, two-dimensional ($d=2$) Lennard-Jones
supercooled liquid. A summary of some of our results has been
published earlier~\cite{prl}.

To help the reader understand our interpretation of the molecular
dynamics data and the questions we pose, we first review the
behavior of a mean-field model of a structural glass
transition~\cite{clumps}. In the mean-field model, particles
interact via a repulsive, two-body potential of the form~\cite{kac}
$V(r) = {\gamma}^d\phi(\gamma r)$, where $r$ is the interparticle
distance and $d$ is the spatial dimension. The range of the
interaction is
$R=\g^{-1}$. In the limit $R \to \infty$, the static properties of
the uniform fluid are described exactly by mean-field
theory~\cite{kac}. The equilibrium structure function $S(k)$ can be
shown to satisfy the Ornstein-Zernicke form, $S(k)=1/[1 +
\beta\rho{\hat\phi}(k)]$, where $\rho$ is the particle density,
$\beta^{-1}=k_BT$, and
${\hat\phi}(k)$ is the Fourier transform of $\phi(x)$. We choose
the step potential,
$\phi(x) = 1$ for $x \leq 1$ and $\phi(x) =0$ for $x > 1$. The
Fourier transform $\hat\phi(k)$ is negative for various values of
$k$. This property of $\hat\phi(k)$ implies that in the mean-field
limit,
$S(k)$ becomes negative in some range of $k$ for sufficiently
large
$\beta
\rho$. Hence for fixed $\rho$, the system has a spinodal
singularity~\cite{grewe} at a temperature $T=T_s$ which is defined
by the condition,
$1+\beta_s\rho{\hat\phi}(k_0)=0$, where $k_0$ is the location of the
minimum of ${\hat\phi}(k)$. At $T=T_s$, $S(k)$ diverges at $k_0$;
for $T>T_s$, $S(k_0)$ diverges as $(T-T_s)^{-\gamma}$ with
the mean-field exponent $\gamma=1$. Note that the spinodal
is the limit of thermodynamic stability of the uniform phase and is
a critical point (for a given density). However, unlike the usual
critical point, the structure function diverges at a nonzero value
of $k$.

For $d=3$ and $\rho=1.95$, the spinodal is at $T_s = 0.705$
($k_B=1$)~\cite{clumps}. Although the singularity is well defined
only in the mean-field limit, simulations must be done for finite
$R$. Monte Carlo simulations for $R=3$ with $\rho=1.95$ yield a
$S(k)$ which has a maximum at $k \neq 0$ that increases rapidly as
$T$ is decreased until $T\approx 0.75$, below which the peak ceases
to increase as $T$ is lowered~\cite{clumps,ramos}. This
behavior is characteristic of a pseudospinodal~\cite{dieter}.

One way of understanding the simulation results for $S(k)$ is
to interpret the spinodal as a singularity of the free
energy as a function of
$T,\rho$. If the interaction range is finite, the singularity is in
the complex $(T,\rho)$ plane as has been found in transfer matrix
studies of long-range ferromagnetic Ising models~\cite{novotny}. The
singularity moves closer to the real axis as the range of
interaction increases. We refer to the complex singularity in finite
range systems as a pseudospinodal. The physical effects of the
pseudospinodal depend on how far it is from the real axis. As $R$
increases, the pseudospinodal better approximates a true singularity
and has measurable effects if $R$ is sufficiently large.

The fact that the mean-field model has a spinodal is unambiguous.
We now discuss the reasons why we can associate the spinodal with a
thermodynamic glass transition. In the Monte Carlo simulation for
fixed $\rho$ we equilibrate the system at a high $T>T_s$ where the
uniform phase is stable and then quench it to $T<T_s$ where the
uniform phase is unstable. After the quench the particles
immediately form clumps with order $\rho R^3$ particles in each
clump~\cite{ramos}. That is, the system breaks up into regions of
high particle density surrounded by regions of low particle
density. The arrangement of the clumps is noncrystalline and their
number depends on the quench history~\cite{clumps,ramos}. The free
energy has been calculated numerically in the mean-field limit and
has many minima corresponding to different numbers of
clumps~\cite{clumps,clejan}. For
$T<T_s$, the system remains trapped in a local free energy minimum.
The minimum the system chooses appears to depend on the quench
history.

The multi-minima free energy and the quench history dependence
suggest that the mean-field model has a metastable glass phase for
$T<T_s$. However, the properties of a glass are associated more
with its dynamical properties. As we will discuss, the glassy
dynamics of the mean-field model is associated with the behavior of
the {\it clumps}; the {\it particles} are not localized in the
metastable glass phase. That is, the dynamical properties
associated with single particle behavior do not show the usual
signature of the approach to a glass. A simple argument based on
the fact that all potential barriers in the mean-field model are
finite~\cite{prl} implies that the self-diffusion coefficient $D$
is nonzero for all $T>0$. Hence, if the observation time is
sufficiently long, the mean square displacement of the particles
increases indefinitely. The argument proceeds as follows. Because
the separation between the clumps is order $R$, the interaction
of the particles in a particular clump with the particles in
neighboring clumps is minimized. Inside a clump, a particle
undergoes a restricted random walk. A particle that attempts to
leave a clump experiences a potential barrier due to the proximity
of other clumps. Such a particle must interact with particles in
the same clump and with particles in at least one other clump at
some time during its possible escape. The upper bound of the
potential barrier is $c\,\g^dR^d$, where the constant
$c$ depends on
$d$. (Recall that $\g^d$ is the strength of the interaction and
$R^d$ is proportional to the number of particles in a clump such
that $\g^dR^d=1$.) The probability of leaving a clump in a Monte
Carlo simulation is bounded from below by $\half e^{-c/k_BT}$
for all $R$ and hence $D>0$. Similar arguments hold for a molecular
dynamics simulation of the same system.

The same argument implies that the mean-field model is ergodic for
all $T>0$ if single particle properties are probed. The ergodic
behavior can be characterized by several fluctuation
metrics~\cite{thirumalai}. The single particle energy fluctuation
metric $\Om_\e(t)$ is given by~\cite{thirumalai}
\beq
\Om_\e(t) = {1\over N} \sum_{i=1}^N \big \lbrack \bar{\e}_i(t) -
\lb\bar{\e}(t)\rb\,\big\rbrack^2, \label{ome}
\eeq
where
\beq
\bar{\e}_i(t) = {1 \over t} \! \int_0^t \! \e_i(t')\, dt'
\label{ebar}
\eeq
is the energy of particle
$i$ averaged over the time interval $t$, and $\lb\bar{\e}(t)\rb =
(1/N)\sum_{i=1}^N
\bar{\e}_i(t)$. The single particle energy $\e_i$ of particle $i$
includes its kinetic energy and one-half of the potential energy
of its interaction with other particles in the system. If the system
is ergodic, $\Om_\e(t) \sim 1/t$ for $t$ sufficiently
large~\cite{thirumalai}. We find that $\Om_\e(t)$ exhibits ergodic
behavior at $T=0.4$, a value of $T<T_s$. However, for $T \le 0.15$,
$1/\Om_\e(t)$ is not proportional to $t$ during our longest runs,
and the measured $D$ is indistinguishable from zero at $T=0.15$.
Given our theoretical argument that $D$ is nonzero for all $T>0$,
the observed behavior of $1/\Om_\e(t)$ for $T< 0.15$ implies only
that the time for a particle to leave a clump is much longer than
our observation time. We interpret this change from ergodic to
nonergodic behavior as an apparent kinetic transition.

How can we reconcile the $T$-dependence of $D$ and $\Om_\e(t)$
with our identification of $T_s$ as the spinodal/glass transition?
The answer lies in the dynamical properties of the {\it clumps}. For
example, the diffusion coefficient of the center of mass of the
clumps, $D_{\rm cl}$, is zero for
$T<T_s$ in the mean-field limit. To understand this
behavior, note that $n_{\rm cl}$, the mean number of particles in a
clump, diverges as $R^d$ as $R \to \infty$. In this
limit, the number of clumps does not change with time and the
mean numbers of particles exiting and entering a clump are
equal. From the central limit theorem, the relative fluctuations of
these quantities goes to zero as $R$ and $n_{\rm cl} \to
\infty$. We conclude that $D_{\rm cl} \leq D/\!\sqrt {n_{\rm cl}}$,
and hence
$D_{\rm cl}=0$ in the mean-field limit. Our simulations of $D_{\rm
cl}$ for finite
$R$ are consistent with this prediction, and also indicate that the
clumps do not see the same local environment, that is, they have
different numbers of nearest neighbor clumps. This difference in the
local environment persists as $R
\to\infty$ because the clumps do not diffuse and cannot sample
different local environments. Hence, the system is nonergodic on a
clump (${\rm mass} \sim R^d$) scale for
$T < T_s$ in the mean-field limit.

In summary, the static properties of the mean-field glass phase are
dominated by localized, long-lived structures (clumps) for $T<T_s$,
even though particles move from clump to clump. The time scale for
the motion of the clumps diverges in the mean-field limit, and the
spinodal/glass transition is seen dynamically only on a clump scale.
However, we observe an {\it apparent} kinetic transition that is
associated with the slow diffusion of the particles and the finite
duration of our runs. The temperature of this apparent transition is
less than $T_s$ and depends on the observation time.

The well characterized behavior of the mean-field glass model
motivates us to ask if similar behavior occurs in systems with more
realistic interactions. In Section~2 we discuss the static properties
of a two component, two-dimensional system of Lennard-Jones
particles and find that the system exhibits the usual properties
associated with other simulations of glassy systems. In Section~3 we
compute the static structure function $S(k)$ and find evidence for a
weak divergence in the height of the diffraction peak as $T$ is
lowered. This divergence is interpreted as evidence for the
influence of a pseudospinodal. In Section~4 we propose a criterion
for clusters of solid-like particles and find evidence of
cluster scaling and an ergodic to nonergodic
transition in the dynamics of the clusters. In Section~5 we
discuss the interpretation of our molecular dynamics results in
terms of a mean-field perspective.

\medskip \noindent {\bf 2. Lennard-Jones model}.

\noindent Because it is easier to identify geometrical structures in
two dimensions than in three, we consider a two-dimensional
system. We also specify that the particles interact via a
Lennard-Jones potential because of the extensive simulations of
supercooled Lennard-Jones liquids. However, a single component,
two-dimensional Lennard-Jones system quickly nucleates to a solid
and is unlikely to form a glass-like state. We follow Wong and
Chester~\cite{wong} who have done a Monte Carlo study of
the quenched states of two component, two-dimensional Lennard-Jones
systems with the goal of choosing the density and the size of the
minority component so that it inhibits nucleation of the majority
component to a solid.

We designate by $V_{ab}$ the interaction between a particle
belonging to species $a$ and a particle belonging to species $b$ and
write $V_{ab} = 4\e[(\s_{ab}/r)^{12} - (\s_{ab}/r)^6 ]$. We follow
ref.~\cite{wong} and choose $\s_{aa} = \s$, $\s_{bb} = 3\s/2$,
and $\s_{ab} = \half(\s_{aa} +\s_{bb}) = 5\s/4$. The energy
parameter $\epsilon$ and the mass $m$ are the same for both
species. The Lennard-Jones
potential is cutoff at $r=2.5\,\sigma$ and shifted so
that the potential goes to zero at the cutoff. We
choose units such that lengths are measured in terms of
$\sigma$, energies in terms of $\epsilon$, and the time in terms of
$\tau = (m\sigma^2/\epsilon)^{1/2}$. For example, the reduced number
density is $\rho^* = \rho \sigma^2$. In the following, we will omit
the asterisk and quote all results in reduced units. Because the
time can be expressed either in terms of the number of time steps or
in terms of $\tau$, we will explicitly specify the units of time.

The majority $a$ component is taken to be 80\% of the total number
of particles $N$. All but one of the simulations are for $N=500$.
This relatively small number of particles was chosen so that long
runs could be made. The duration of the run for each temperature was
$100\,000\,
\tau$ unless otherwise noted. The molecular dynamics simulations were
performed with periodic boundary conditions at constant energy and
volume~\cite{gt} at a density of
$\rho = 0.952$, corresponding to $L=22.92$, where $L$ is the linear
dimension of the simulation cell. We used the velocity form of the
Verlet algorithm~\cite{gt} with a time step of
$\Delta t = 0.005\, \tau$ for all temperatures except for $T=5.5$
for which we choose $\Delta t = 0.0025\, \tau$ to achieve
reasonable energy conservation. The mean
temperatures and pressures of our eight different runs are shown in
Table~\ref{table:runs}. As a typical example of the statistical
accuracy of these mean values, we note that the run at $T=3.1$
corresponds to a mean temperature of 3.15 over the first
$50\,000\,\tau$ and 3.13 over the second $50\,000\,\tau$. However
for $T=2.25$, the mean temperatures were 2.21 and 2.29
respectively, and it is possible that the lowest temperature run
has not reached thermal equilibrium.

Because our results are for constant density rather than constant
pressure, the values of various quantities to be discussed here will
differ slightly from our earlier results cited in ref.~\cite{prl}
which were for a pressure of
$P \approx 70$ and different densities. The initial configuration of
each run at a particular temperature was the final configuration
obtained previously~\cite{prl,melcuk}. In some cases the positions
had to be rescaled to obtain the desired density. Unless otherwise
stated, the system was allowed to equilibrate for at least
$50\,000\,
\tau$ and then data was collected for $100\,000\, \tau$. This
duration is one to two orders of magnitude
longer than reported previously~\cite{prl} and is equivalent to
runs of approximately $2 \times 10^{-7}\, {\rm s}$ for liquid Argon.

We first compute the radial distribution function $g(r)$ and the
self-diffusion coefficient $D$ and show that they exhibit behavior
similar to that found in other simulations of supercooled liquids.
All of the following results are for the majority particles only
unless otherwise noted. The positions of the particles were saved
every $5\, \tau$ and $g(r)$ was computed in a separate program. In
Fig.~\ref{fig:gr} we show $g(r)$ for $T=2.7$. Note the split second
peak, a possible signature of a glassy state~\cite{wong}; the
split second peak of $g(r)$ is not observed for $T>3.1$. The
increase in the height of the first peak of $g(r)$ as the
temperature is lowered (see Table~\ref{table:runs}) indicates the
growth of short-range order. We discuss the temperature dependence
of the enthalpy and the constant pressure heat capacity in
Appendix~A.

The single particle self-diffusion coefficient $D$ is related to
the mean square displacement $R^2(t)$ by $\lim_{t \to \infty}
R^2(t) = 2d Dt$. An alternative way of determining $D$ is from the
velocity fluctuation metric $\Om_v(t)$ defined as~\cite{thirumalai}
\beq
\Om_v(t) = {1 \over N} \sum_{i=1}^N {1 \over d} \sum_{\a=1}^d
\bigl [\bar{v}_{i,\a}(t) - \lb \bar{v}_\a(t) \rb \bigr]^2 ,
\label{omv}
\eeq
where $v_{i,\a}$ is the $\a$-component of the velocity
$\vv_i$ of particle $i$. The time average, $\bar{v}_{i,\a}(t)$, of
$v_{i,\a}$ is defined as in (\ref{ebar}) and the quantity $\lb
\bar{v}_\a(t)\rb$ is the average of
$\bar{v}_{i,\a}(t)$ over all particles. In our simulations the total
momentum of the system is equal to zero, and hence
$\lb \bar{v}_\a(t)\rb=0$. Because $\int_0^t \vv_i(t') \, dt' =
\rv_i(t) - \rv_i(0)$, we can express $\Om_v(t)$ as 
\beq
\Om_v(t) = R^2(t)/2t^2 .
\eeq
If the particles are undergoing diffusion, we have
\beq
\Om_v(t) \to {2 D \over t} . \label{omDt}
\eeq
The relation (\ref{omDt}) was also used to determine $D$.

For each value of $T$, the quantity $|\rv_i(t+t_0)-\rv_i(t_0)|^2$
was computed in a separate program by grouping the particle
coordinate data into blocks of duration $500\,\tau$ and averaging
over all possible choices of the origin $t_0$ and combinations of
time differences
$t \leq 250$ for a particular block. The results were then averaged
over the 200 blocks.
The velocity metric was computed at each time step in the main
molecular dynamics program and the results also were averaged over
200 time origins. The typical time-dependence of
$\Om_v(0)/\Om_v(t)$ is shown in Fig.~\ref{fig:omv} for $T=2.25$.
Note that at this temperature
$1/\Om_v(t)$ is not linear for $t < 200\,
\tau$, but becomes linear for longer observation times.

Our results for $D$ obtained from $R^2(t)$ and $1/\Om_v(t)$ are
summarized in Table~\ref{table:D}. The difference in the values of
$D$ determined by the two methods is a measure of the error in the
determination of $D$. In the similar three-dimensional system
studied in ref.~\cite{clejan}, $D(T)$ was fit to the Vogel-Fulcher
form
\beq
D(T) = A\, e^{-B/(T-T_0)} , \label{dvf}
\eeq
and we look for similar behavior. From the semilog plot of $D$ versus
$1/T$ shown in Fig.~\ref{fig:dt}, it is easy to verify that the best
fits are for $T_0 > 0$. The best fit for all eight temperatures
in Table~\ref{table:D} occurs for $T_0 \approx 1.8$ and $T_0
\approx 1.7$ using the values of $D$ obtained from $R^2(t)$ and
$1/\Om_v(t)$, respectively. However, as shown in
Table~\ref{table:fitD} the results for $T_0$ are sensitive to the
number of data points which are included in the least squares fits.
The range of reasonable fits illustrates
the difficulty of obtaining meaningful values of
$T_0$. We conclude that the temperature-dependence of $D(T)$ is
consistent with the Vogel-Fulcher form with $T_0$ in the range $1
\leq T_0
\leq 2$. Note that the Vogel-Fulcher form of $D(T)$ implies that
the system loses ergodicity at $T = T_0$.

We also calculate the single particle energy fluctuation metric
$\Om_\e(t)$ (see (\ref{ome})) to see if it exhibits ergodic
behavior. If the system is ergodic, we expect that~\cite{thirumalai}
\beq
\Om_\e(t) \to 2 \tau_\e/t . \qquad (t>>1) \label{ometau}
\eeq
We interpret $\tau_\e$ as the energy mixing time. The energy
fluctuation metric was computed ``on the fly'' in the same way as
$\Omega_v(t)$, but for a duration of $50\,000\, \tau$ rather than
$100\,000\, \tau$. We observed that $1/\Om_\e(t)$ becomes
approximately linear even at the lowest temperature $T=2.25$,
indicating that the system is ergodic according to this measure.
Our results for the $T$-dependence of
$\tau_\e(T)$ based on (\ref{ometau}) are summarized in
Table~\ref{table:times} and are plotted in Fig.~\ref{fig:emetric}.
Note that the largest mixing time is order $10\,000\,\tau$. For
comparison, we also show in Table~\ref{table:times} the values of
$\tau_D=\sigma^2/(4D)$, the mean time it takes a particle to
traverse the distance $\sigma$. As summarized in
Table~\ref{table:fitD}, the fits of
$\tau_\e(T)$ to the Vogel-Fulcher form $\tau_\e(T) = A\,
e^{-B/(T-T_\e)}$ in different temperature intervals also yields a
range of values for the parameter
$T_\e$. We conclude that the
temperature-dependence of $\tau_\e$ is consistent with the
Vogel-Fulcher form with a value of $T_\e$ consistent with the value
of $T_0$ determined from the self-diffusion coefficient. 

\medskip \noindent {\bf 3. Evidence of pseudospinodal behavior}

\noindent Now that we have established that the two component,
two-dimensional supercooled Lennard-Jones system has properties
similar to those observed in other simulations of deeply supercooled
systems, we explore how the mean-field interpretation discussed in
Section~1 is applicable to the present Lennard-Jones system. If the
Lennard-Jones system exhibits pseudospinodal effects, we should find
behavior analogous to that observed in the mean-field glass
model~\cite{clumps} and in Ising models with long, but finite range
interactions~\cite{dieter}. In these systems, the static structure
function $S(k)$ appears to diverge at a nonzero value of $k$ if its
behavior is extrapolated from high $T$ or small magnetic field, but
the extrapolated singularity is not observed if measurements are
made too close to the apparent singularity.

We compute $S(k)$ (for the majority particles) from the saved
particle positions using the relation
\beq
S(k) = {1 \over N} \bigl |\sum_{i=1}^N e^{i \kv \cdot \rv_i}
\bigr |^2 .
\eeq
Angular averages were computed by considering all possible
$k$-vectors. We calculated $S(k)$ for the same configurations as
were used for obtaining $g(r)$. The $k$-dependence of $S(k)$ at
$T=2.7$ is plotted in Fig.~\ref{fig:sk}. Because the number of
particles is fixed, we expect that $S(k) \to 1$ as $k \to 0$. The
important feature of $S(k)$ is the diffraction peak at $k=k_0
\approx 7.15$. We interpret the height of this peak as a
$k$-dependent susceptibility $\chi(k_0,T)$, and the width $w(k_0,T)$
as an inverse length proportional to the size of the correlated
regions in the liquid. These quantities were extracted from $S(k)$
by fitting the region around $k=k_0$ to the four parameter form,
$S(k) = a + b/[(k-k_0)^2 + c]$.

Our results for $\chi(k_0,T)$ and $w(k_0,T)$ are listed in
Table~\ref{table:runs} and plotted in Fig.~\ref{fig:chi}. Note that
$\chi(k_0,T)$ increases by a factor of approximately 1.6 and
$w(k_0,T)$ decreases by a factor of approximately 2.0 as $T$ is
lowered from 5.5 to 2.5. Because this range of $T$ is limited, we
can fit $\chi(k_0,T)$ and $w(k_0,T)$ to a variety of functional
forms. Given the divergent behavior of the first peak of $S(k)$ in
the mean-field model discussed in Section~1, we look for fits to the
functional forms, $\chi(k_0,T) \sim (T-T_s)^{-\gamma}$ and
$w(k_0,T) \sim (T-T_s)^\nu$, where $T_s$, $\gamma$, and $\nu$ are
fit parameters. We find that the best fit to the assumed power law
form is $\chi(k_0,T)
\sim (T - 2.6)^{-0.16}$ and $w(k_0,T)
\sim (T - 2.6)^{0.29}$ if results in the interval $2.7 \leq T
\leq 4.6$ are included. A least squares fit in the interval $3.1
\leq 5.5$ yields $\chi(k_0,T) \sim (T-3.0)^{-0.08}$ and $w(k_0,T)
\sim (T - 3.0)^{0.19}$. Given the limited range of
$T$, these power laws fits are justified {\it only} in the context
of our rigorous results for $S(k)$ in the mean-field
model~\cite{clumps} for which
$\chi(k_0,T)
\sim (T-T_s)^{-1}$. The power law fits suggest that the increase of
the height and the decrease in the width of the first peak of $S(k)$
is influenced by a weak singularity at $T = T_s$ with $T_s$ in the
range $2.6 \leq T_s \leq 3.0$. As expected, no evidence of a
singularity is found if we fit the results for $\chi(k_0,T)$ in the
interval $2.25 \leq T \leq 5.5$.

\medskip \noindent {\bf 4. Cluster scaling and lifetime}.

\noindent Given the preliminary pseudospinodal interpretation that
we presented in Section~3, we seek other evidence for the existence
of a pseudospinodal and mean-field behavior. For temperatures near
the pseudospinodal the system should show signs of an instability and
the system should partially phase separate. That is, we should see
regions where the majority particles dominate and locally order.
For this reason we look for long-lived structures whose constituent
particles remain in close proximity to each other over extended
times at sufficiently low $T$. Because of the strong repulsive
Lennard-Jones interaction near the origin, these structures will not
be identical to the clumps found in the mean-field model, but
instead should be caused by a cage effect. A visual examination of
the configurations shows evidence of a partial phase separation in
which a significant fraction of the majority particles form
clusters of triangular-like structures which become better defined
as the temperature is decreased. Qualitatively, the lifetime of
these visual clusters in the range of temperatures of interest
appears to be much longer than the period of oscillation of the
particles about their mean position in the clusters.

To find these structures we seek to define a ``solid-like'' particle
such that clusters of these particles have the above qualitative
properties. However, unlike the Ising model near the critical
point~\cite{coniglio} and near the spinodal~\cite{klein}, there is
no theoretical definition of clusters in a dense liquid, and we
have to rely on our physical intuition to define them. We will
assume that a cluster consists of a group of solid-like particles
and that if two solid-like particles are near neighbors, they
belong to the same cluster. Ideally, we would like to introduce a
physical property that exhibits bimodal behavior with one peak
corresponding to solid-like particles. However, all of the physical
quantities we measured exhibit a single peak, and hence we will
need to introduce a cutoff to distinguish solid-like and
non-solid-like particles. Nevertheless, we will find that the
properties of the resultant clusters do not depend strongly on the
choice of cutoff.

Because we are interested in the local environment of the particles,
it is natural to do a Voronoi construction~\cite{voronoi} and
determine the Voronoi polygon of each particle and its Voronoi
neighbors. Quantities associated with the Voronoi construction
include the distribution of the number of edges of the Voronoi
polygons, the distribution of the area of the polygons, and the
distribution of the length of the sides. For our two-dimensional
system, the mean number of edges of the Voronoi polygons is six and
the distribution of the number of edges is
temperature-independent. This independence is expected
because the Voronoi construction in two dimensions is insensitive to
thermal fluctuations~\cite{melcuk}.

To quantify the temperature-dependent changes in the
distribution of lengths of the sides of the Voronoi polygons with
six sides, we introduce the standard deviation
$\s_\ell$ of the edge length of a particular particle
as~\cite{medvedev}
\beq
\sigma_\ell^2 = \lb \ell^2 \rb - \lb \ell \rb^2 ,
\eeq
where $\lb f(\ell) \rb = (1/6)\sum_{\a=1}^6 f(\ell_\a)$ and
$\ell_\a$ is the length of edge $\a$ of the hexagon of interest. If
a particle were in a perfect triangular environment, then
$\sigma_\ell = 0$. In Fig.~\ref{fig:ph} we show the estimated
probability density $P(h)$ of the quantity
$h=\sigma_\ell^2/\lb \ell \rb^2$, a measure of the ``hexagonality''
of the polygons. At $T=5.5$, the Voronoi polygons are irregular,
giving rise to a high percentage of particles with $h > 0.1$. At
$T=2.7$, the most probable value of $h$ is at $h=0.014$, and many
more particles have regular Voronoi polygons.

In the following, we define a majority particle to be solid-like if
it has six Voronoi neighbors and if the condition $h \leq 0.1$ is
satisfied. The choice of the cutoff parameter is not crucial and
the qualitative properties of the clusters are {\it independent} of
the cutoffs over a wide range of values~\cite{melcuk}. A typical
configuration of the system at $T=3.3$ is shown in
Fig.~\ref{fig:config}.

Given our criteria for solid-like particles and our definition of
clusters, we can determine the properties of the clusters. A
separate program using a Voronoi construction and a Hoshen-Kopelman
cluster labeling algorithm was used to determine the clusters.
Because the height of the maximum of $S(k)$ increases and the width
of the peak decreases as $T$ is lowered, we expect that the mean
size of the clusters grows as $T$ is lowered until their growth
becomes ``frustrated'' by the presence of the larger minority
component and by the different orientations of the other clusters.
(The size of a cluster is its number of particles.) The behavior of
the cluster size distribution
$n_s$ as a function of the size $s$ is plotted in
Fig.~\ref{fig:ns3.3} for $T = 3.3$ and $N=500$. The cluster
distribution is normalized by the number of (majority) particles so
that $n_s$ is
the probability that a particle belongs to a cluster of size $s$.
The results for $n_s$ are averaged over $100\,000\, \tau$. We
expect that the presence of a pseudospinodal leads to power law
scaling of the clusters if
$T$ is close, but not too close to the pseudospinodal, that is, we
expect
$n_s
\sim s^{-x}$ for a range of values of $s$. The lack of a true
spinodal should lead to a cutoff or lack of scaling for large
clusters or for temperatures far from the pseudospinodal. From
the log-log plot of $n_s$ versus $s$ in Fig.~\ref{fig:ns3.3} we see
that the $s$-dependence of $n_s$ is consistent with a power law
dependence with an exponent of $x \approx 1.8$ over approximately
one decade of $s$.

The behavior of $n_s$ for lower temperatures is qualitatively
different. For example, compare the behavior of
$n_s$ averaged over the first $50\,000\, \tau$ of the run at
$T=3.1$ to the plot of $n_s$ averaged over the second $50\,000\,
\tau$ of the run (see Fig.~\ref{fig:ns3.1}). The difference in the
cluster size distribution for larger values of $s$
indicates that the lifetime of the clusters at $T=3.1$ is the same
order of magnitude as the duration of our runs. The fact that larger
clusters require a longer time to reach equilibrium than smaller
clusters has been found previously in temperature quenches of Ising
models near the spinodal~\cite{monette}. Our simulation results for
$n_s$ for lower values of $T$ show similar behavior. We will
discuss other estimates of the lifetime of the clusters in the
following.

The behavior of $n_s$ for $T=5.5$ is shown in Fig.~\ref{fig:ns3.7}.
No evidence for simple power law behavior is observed, but the
$s$-dependence of $n_s$ is consistent with a fit to the
assumed form, $n_s \propto s^{-x} e^{-s/m_s}$ with $x =
1.1$, and $m_s=13$. Similar fits can be made for $T=4.7$ with
$x=1.2$, and $m_s=21$, and $T=3.7$ with
$x = 1.4$, and $m_s = 42$. We note that the effective value of the
power law exponent $x$ and the cutoff parameter $m_s$ increases as
the temperature is decreased, suggesting that a weak singularity is
being approached.
 
Because our results for $n_s$ for $N=500$ might be affected by
finite size effects, we did a run for $N=20\,000$ particles at
$T=2.7$ for a time of $30\,000\, \tau$. The system was run for
$15\,000\, \tau$ before data was collected. A log-log plot of $n_s$
versus $s$ is shown in Fig.~\ref{fig:ns20000}. It is clear that
the values of $n_s$ for larger $s$ are not in equilibrium. However,
a power-law fit of $n_s$ in the range $4 \leq s \leq 61$ yields a
slope of $x=1.75$, a value of $x$ consistent with the value
estimated from our results for $N=500$ at $T=3.3$.

We expect that if the clusters are important near the glass
transition, their lifetime should increase as the glass transition
is approached. As indicated in Fig.~\ref{fig:ns3.1}, we know
qualitatively that the lifetime of the larger clusters becomes very
long as the temperature is reduced. We introduce a measure of the
cluster lifetime by measuring the time-dependence of a metric
associated with the solid-like particles. We divide the simulation
cell into boxes and compute the number of solid-like particles in
each box. The idea is to find if the time averaged number of
solid-like particles in each box becomes the same when the time
$t>>1$. We take $n_\a$ to be the number of solid-like particles in
box
$\alpha$ and compute the cluster fluctuation metric $\Om_{\rm
cl}(t)$ defined as
\beq
\Om_{\rm cl}(t) = {1 \over N_b} \sum_{\a=1}^{N_b}
\bigl [\bar{n}_\a(t) -
\lb \bar{n}(t)\rb \bigr]^2 . \label{omcl}
\eeq
In (\ref{omcl}) $\bar{n}_\a(t)$ is the mean
number of solid-like particles in box $\a$ at time $t$, and
$\lb\bar{n}(t)\rb$ is the mean occupancy averaged over all $N_b$
boxes. For our runs we divided the system into $5
\times 5$ boxes.
A linear increase in $1/\Om_{\rm cl}(t)$ defines the cluster mixing
time
$\tau_{\rm cl}$ as in (\ref{ometau}).

At $T=5.5$, $\Om_{\rm cl}(t)$ exhibits ergodic behavior with
$\tau_{\rm cl} \approx 500\, \tau$ (see Fig.~\ref{fig:omcl}a). Our
estimates for $\tau_{\rm cl}$ for our runs are summarized in
Table~\ref{table:times}. Note that for
$T=3.1$ (see Fig.~\ref{fig:omcl}b), $\tau_{\rm cl}$ is order $5
\times 10^4\,\tau$, a time which is comparable to the duration of
runs. For $T<3.1$, our estimates of
$\tau_{\rm cl}$ are longer than the duration of our runs and are not
meaningful. We interpret the time $\tau_{\rm cl}$ as an estimate of
the lifetime of the clusters. Note that at
$T=3.1$, the times associated with the cluster lifetime and the
motion of single particles differ by a factor of $10^3$. The
$T$-dependence of $\tau_{\rm cl}$ is best approximated by a
Vogel-Fulcher form $\tau_{\rm cl} \sim e^{C/(T - T_{\rm cl})}$,
where $T_{\rm cl}$ is the extrapolated temperature at which the
cluster lifetime would become infinite. Although our estimates for
$\tau_{\rm cl}$ are only qualitative, the estimated value of
$T_{\rm cl}$ found by considering the values of $\tau_{\rm cl}$ in
the range $3.1 \leq T \leq 5.5$ yields a reasonable fit with
$T_{\rm cl} \approx 1.6$ (see Fig.~\ref{fig:tau_cl}). This estimate
of $T_{\rm cl}$ does not vary much if the results at $T=3.3$ and
$T=3.7$ are omitted (see Table~\ref{table:fitD}). 

Another single particle decorrelation time can be extracted from
$n_b(t)$, the number of unbroken Voronoi bonds remaining after a
time $t$. At
$t=0$, only Voronoi bonds between small particles that have exactly
six small neighbors are counted. If at a time $t$ later, there is
no longer a bond which joins the same pair of particles, the number
of bonds is reduced, and we do not consider this pair of particles
again. We find that
$\lb n_b(t)n_b(0) \rb \sim e^{-t/\tau_b}$. Because a bond is broken
every time a particle changes neighbors, we expect that $\tau_b \sim
\tau_D$. We computed $\tau_b$ for only a few temperatures and found
that $\tau_b$ is comparable to $\tau_D$.

Given the very long lifetime of the clusters for $T \leq 3.1$, it is
difficult to make estimates of the errors associated with various
quantities. For example, even though we made long runs at low
temperatures, we found only one statistically independent
configuration as far as the clusters are concerned. For this
reason, quantities which are measures of the structure of the
system, such as
$S(k)$, are probably not adequately sampled for $T \leq 3.1$. And
although our measures of single particle properties such as the
velocity metric show the system to be ergodic at this level, the
values of
$D$ at low $T$ might also be inadequately sampled because the motion
of the particles is influenced by the presence of the long-lived
clusters.

It is not clear from our results and various fits whether we can
identify one or two temperatures which can be associated with a
glass transition. Our estimates for the temperature $T_0$ at which
the self-diffusion coefficient vanishes and measures of the single
particle properties become nonergodic are in the range
$1 \leq T_0 \leq 2$. In comparison, our estimates for the
temperature $T_s$ at which the $k$-dependent susceptibility
$\chi(k_0)$ and the inverse width $w^{-1}(k_0,T)$ would diverge if a
spinodal were present are in the range $2.6 \leq T_s \leq 3.0$. The
cluster distribution exhibits power law scaling at a temperature
which is close, but not too close, to our estimate of $T_s$. At
$T \approx 3.1$, the lifetime of our clusters is already comparable
to the duration of our runs. Nevertheless, if we extrapolate the
cluster lifetime to lower temperatures by fitting the cluster
lifetime to a Vogel-Fulcher form, we find that the cluster lifetime
becomes infinite at $T_{\rm cl} \approx 1.6$, a temperature
consistent with our estimate of $T_0$.

Given the small size of our system for all but one of our runs and
the limited number of temperatures available, it is difficult to
make quantitative conclusions in spite of the relatively long
duration of our runs. Moreover, as we have emphasized, our
interpretation of our molecular dynamics data is justified only in
the context of our rigorous results for the mean-field model of a
structural glass discussed in Section~1.

\medskip \noindent {\bf 5. Mean-field interpretation and summary.}

\noindent On the basis of our molecular dynamics simulations, we
can conclude that the deeply supercooled, two component,
two-dimensional Lennard-Jones system is heterogeneous. The
heterogeneity can be characterized both dynamically and statically,
that is, there are long-lived spatially correlated regions of all
sizes up to a cutoff. This qualitative conclusion is consistent
with the results of rotational diffusion experiments on probe
molecules in supercooled o-terphenyl~\cite{ediger}, which indicate
that the dynamics in supercooled o-terphenyl is spatially
heterogeneous, and with the results of other
simulations~\cite{mountain2}.

An important question is the appropriate theoretical understanding
of the origin of the spatial heterogeneity. In a future publication
we will give a mean-field argument for the origin of the scaling
behavior of the cluster size distribution in terms of an arrested
nucleation picture~\cite{kleinpc}.

We have found indirect evidence for mean-field behavior and the
influence of a pseudospinodal. As discussed in Section~2, the
height of the diffraction peak of the static structure function
$S(k)$ for the mean-field structural glass model exhibits a true
divergence~\cite{grewe} in the limit $R \to
\infty$. For finite range $R$, the peak of $S(k)$ appears to
diverge if it is measured for values of the temperature $T$ which
are close, but not too close, to the apparent singularity and the
data is extrapolated to lower $T$. However, a singularity in the
peak of
$S(k)$ is not observed if measurements are made too close to the
apparent singularity. This behavior of
$S(k)$ is characteristic of a pseudospinodal. Is there a spinodal
in the Lennard-Jones system? The answer is no, because the range of
the Lennard-Jones potential is finite. However, we found that the
height and the inverse width of the diffraction peak of $S(k)$
exhibits a weak power law divergence if their behavior is
extrapolated from high $T$. This behavior is consistent with a
pseudospinodal interpretation.

We expect that the Lennard-Jones system would be better described by
mean-field theory as the density is increased, because the
number of interactions each particle experiences increases. At
present, we do not know how to calculate the effects of the
pseudospinodal in dense Lennard-Jones systems~\cite{lovett}, and we
need to rely on simulations. Several other measurements suggest that
the deeply supercooled, dense Lennard-Jones liquid can be described
at least qualitatively by a mean-field picture. Glaser and
Clark~\cite{clark} found cluster scaling in a
simulation of a two-dimensional Lennard-Jones system near the
freezing transition. On the basis of a mean-field theory, it has
been predicted that near the liquid-solid
spinodal, the nucleating droplets are fractal-like rather than
compact objects~\cite{unger,leyvraz}. This effect has been observed
in simulations of a three-dimensional Lennard-Jones
system~\cite{yang}. A third result consistent with a mean-field
interpretation is that the measured value of the fractal dimension
of the structures formed in a single component two-dimensional
Lennard-Jones system undergoing spinodal decomposition is
consistent with that predicted by mean-field
theory~\cite{klein,desai}. These results, together with the results
reported here, suggest that a mean-field interpretation is
applicable to dense Lennard-Jones systems under the proper
conditions.

We note that although the clumps in the mean-field glass model and
the solid-like clusters in the Lennard-Jones system have some
properties in common, for example, their long lifetime, the clumps
are not directly analogous to the clusters. As discussed in
ref.~\cite{lou}, the clump size distribution is a Gaussian in
contrast to the power law distribution that we found for the
solid-like clusters. Although a theoretical definition of the
clusters in the mean-field glass model does not exist, we can
follow a similar approach and assume that a cluster in the latter
system is a group of clumps such that each clump is in a triangular
environment (in two dimensions). A preliminary
investigation~\cite{lou} of such a criterion for a cluster of clumps
yields clusters which have a power-law distribution near the
spinodal consistent with our results for the Lennard-Jones system.

Our argument for the association of the glass transition with an
underlying thermodynamic transition is consistent with the recent
interpretation by Nagel and coworkers~\cite{dixon} of the frequency
dependence of $\e''(\om)$, the imaginary part of the dielectric
susceptibility, in organic glass forming liquids. Nagel and
coworkers~\cite{dixon} have fitted $\e''(\om)$ to a single scaling
curve over 13 decades of frequency for a wide range of $T$ and for
many glass formers. If the temperature-dependence of the high
frequency, power law behavior of $\e''(\om)$ is extrapolated to
lower frequencies, they found that the static susceptibility
diverges at a temperature
$T_\s$ with the corresponding mean-field exponent~\cite{menon}. The
magnetic susceptibility in a dipolar-coupled Ising spin glasses is
found to have similar behavior. Given that these experiments have
been done on fragile glass forming liquids
with long molecules and on dipolar magnets, a mean-field
interpretation of these results is consistent with our picture of
the glass transition. That is, these systems are well approximated
by mean-field models due to the large number of simultaneous
interactions which each molecule has with its
neighbors~\cite{binder}.

Our interpretation of our simulations in
terms of an underlying thermodynamic transition in Lennard-Jones
systems is consistent with the results of recent laboratory
experiments. However, the interpretation of the single particle
behavior in terms of a distinct kinetic transition is more open to
question. Although such an interpretation is rigorous for our
mean-field model of a structural glass, Nagel and
coworkers~\cite{menon} have interpreted their experimental results
in terms of a {\it single} glass transition temperature. That is,
the temperature
$T_\s$ at which the static susceptibility is extrapolated to diverge
is the same temperature at which the self-diffusion coefficient is
extrapolated to vanish. In contrast, we find two distinct
temperatures in the mean-field glass model and also can interpret
our simulation results of the Lennard-Jones system in terms of two
temperatures. We note that in the mean-field limit the number of
particles in each clump is infinite, and the system would remain
indefinitely in a local free energy minimum as determined by the
number and location of the clumps. The duration of our simulations
of the Lennard-Jones system is sufficiently short ($\approx
10^{-7}\, {\rm s}$) that at low temperatures the clusters
do not diffuse and particles in the interior of the clusters are
trapped. Moreover, the mean cluster lifetime near the glass
transition is estimated to be an order of magnitude longer than our
longest runs. This picture of static clusters is consistent with
the mean-field model, but might not be appropriate for
experimental time scales. Hence, it is possible that if we were
able to run for longer times, we would find that the extrapolated
temperature at which the self-diffusion coefficient vanishes and
the extrapolated temperature of the underlying thermodynamic glass
transition would approach each other.

Other workers have also interpreted the behavior of supercooled
liquids near the glass transition in terms of clusters. Kivelson and
coworkers~\cite{kivelson} have proposed a thermodynamic theory of
supercooled liquids based on the assumed existence of a ``narrowly
avoided'' thermodynamic phase transition. The avoided transition is
attributed to the existence of strain. In contrast to this
transition, the spinodal transition always occurs below the
first-order transition temperature. The reason that the spinodal
transition is ``avoided'' in our interpretation is due to the fact
that the system is not really mean-field.

On the basis of our results for the temperature dependence of
$w(k,T)$, the width of the first peak of $S(k)$, we can conclude
that there is an increasing length scale which is associated with
the clusters as the temperature is decreased. In our interpretation
this increasing length scale is due to the effects of the
pseudospinodal. This length scale would diverge if a spinodal were
really present. The relation of this increasing length scale to the
increasing maximum length scale for propagating transverse current
correlations observed by Mountain~\cite{mountain} requires further
study. In contrast, an increasing correlation length was not
observed in a simulation~\cite{grest} of the translational and
orientational correlation functions of a two-component three
dimensional Lennard-Jones system.

The effects of the pseudospinodal and the incipient
thermodynamic glass transition will be more or less apparent
depending on the interaction range, the details of the interaction,
and the spatial dimension~\cite{dieter}. We do not expect to find
spinodal-like effects in all supercooled liquids. These
considerations suggest that there is a class of materials for which
the observed glass transition is associated with a pseudospinodal
and an incipient thermodynamic transition, and other materials for
which the observed glass transition might not be associated with
such effects.

Based on our Monte Carlo and theoretical studies of a mean-field
model and our molecular dynamics results for a two-dimensional
Lennard-Jones system, we suggest that the latter is in the class of
systems whose behavior can be attributed to an incipient
thermodynamic instability (the pseudospinodal). We emphasize that a
true thermodynamic glass transition does not exist in the
Lennard-Jones system, even though the pseudospinodal has measurable
effects including increasing length and time scales as the
pseudospinodal is approached. In addition to this
glass/pseudospinodal transition, there is a temperature (for fixed
density) which can be interpreted as a kinetic transition below
which the self-diffusion coefficient is not measurable during our
observation time.

We are presently simulating much larger Lennard-Jones systems in two
dimensions to obtain better statistics for the clusters over a
wider range of sizes and over a range of temperatures above the
glass transition. If our mean-field interpretation is correct, we
should be able to observe similar behavior in three dimensions
where mean-field behavior should be even more apparent. However,
the identification of the clusters in three dimensions is not
straightforward because of the existence of a variety of possible
local symmetries which are more affected by thermal fluctuations
than in two dimensions.

\vskip 6pt \noindent {\bf Acknowledgments}. This research was
supported in part by the NSF DMR-9632898. \nobreak{Acknowledgment}
is made to the donors of the Petroleum Research Foundation,
administrated by the American Chemical Society, for partial support
of the research at Clark University. This research was based in
part on the Ph.D.\ thesis of Andrew Mel'cuk at Clark University. We
thank Francis J.\ Alexander, Richard Brower, Louis Colonna-Romano,
and Raphael A.\ Ramos for useful conversations.

\medskip \noindent {\bf Appendix A. The enthalpy.}

\noindent Many computer simulations of glasses show that the heat
capacity $C_P$ has a maximum in the vicinity of the glass
transition. For example, Wahnstr\"om~\cite{wahnstrom} has computed
the temperature-dependence of the energy of a two-component, three
dimensional Lennard-Jones system and has found that the slope
changes at a temperature where the dynamical anomalies are most
pronounced. In the following, we discuss our measured values of the
temperature dependence of the enthalpy $H$.

We measured $H$ using constant pressure molecular
dynamics~\cite{andersen} for $N=500$ particles. The system was
equilibrated at
$T
\approx 10$ and a series of measurements of $H$ were performed at
progressively lower temperatures spaced approximately 0.02 apart for
the higher values of $T$ and approximately 0.01 apart at the lower
values of $T$. At each value of
$T$, the system was equilibrated for $250\, \tau$ and $H$ was
averaged over the following $250\, \tau$. These runs are relatively
short in comparison to the runs reported in the main text.

Our results at $P=70$ for $H$ and $C_P$ are shown in
Fig.~\ref{fig:appen}. A careful inspection of $H(T)$ shows that its
slope changes as a function of $T$. Note that $C_P(T)$ increases as
$T$ is lowered, reaches a maximum at $T \approx 4$ and decreases as
$T$ is lowered further. The slope of $H(T)$ was computed as each
value of
$T=T_i$ from the numerical derivative,
$[H(T_{i+1}) - H(T_i)]/[T_{i+1} - T_i]$. The values of
$C_P(T=T_i)$ shown in Fig.~\ref{fig:appen} were computed by doing a
least squares fit to 15 successive slopes in the interval $T_{i+7} -
T_{i-7}$. If we consider less than 15
points, the results for
$C_P(T)$ were too noisy. The results for more than 15 points tended
to smooth the peak in $C_P$.

We note that the spinodal singularity in the mean-field glass model
is at $k=k_0 \neq 0$, and hence we do not expect thermodynamic
quantities such as $C_P$ to exhibit a divergence at the spinodal in
the mean-field limit. However, for a system that does not exhibit a
true spinodal such as the present Lennard-Jones system, it is
possible that the coupling between the $k=k_0$ and $k=0$ modes might
lead to a maximum in quantities such as $C_P$ near the
pseudospinodal.

\bigskip
\begin{table}
\caption{Values of the mean temperature $T$, the mean pressure
$P$, the height of the first peak of the radial distribution
function $g(r)$, and the height
$\chi(k_0)$ and width $w(k_0)$ of the diffraction peak of the
static structure function $S(k)$ averaged over runs of
$100\,000\,
\tau$ each. The mean pressure was computed from the virial. The
width of the first peak of $g(r)$ does not appear to change with
$T$ and is not listed. The density
$\rho$ is fixed at
$\rho = 0.952$. As stated in the text, the lowest temperature run
is probably not in complete thermal equilibrium.}

\label{table:runs}
\begin{tabular}{|rrrrr|}
$T$ & $P$ & $g(r_{\rm max})$ & $\chi(k_0)$ & $w(k_0)$ \\
\tableline
5.5  & 91.1 & 4.30 & 2.59 & 1.11 \\
4.6  & 85.3 & 4.54 & 2.70 & 0.95 \\
3.7  & 77.8 & 4.91 & 2.87 & 0.87 \\
3.3  & 74.9 & 5.15 & 3.10 & 0.76 \\
3.1  & 72.9 & 5.33 & 3.49 & 0.63 \\
2.7  & 69.0 & 5.68 & 4.28 & 0.45 \\
2.5  & 67.3 & 5.86 & 3.94 & 0.55 \\
2.25 & 65.0 & 6.07 & 3.54 & 0.54 \\
\end{tabular}
\end{table}

\begin{table}
\caption{Summary of results for the self-diffusion coefficient
$D$ as a function of
temperature $T$ at constant density $\rho = 0.952$. The second
column represents the estimates of
$D$ determined from the slope of
$R^2(t)$, and the third column gives the estimates of
$D$ from the slope of $1/\Om_v(t)$ (see (\ref{omDt})).}
\label{table:D}
\begin{tabular}{|rrr|}
$T$ & $D$ [from $R^2(t)$] & $D$ [from $1/\Om_v(t)$] \\
\tableline
5.5  & $4.4 \times 10^{-2}$ & $4.8 \times 10^{-2}$ \\
4.6  & $3.0 \times 10^{-2}$ & $3.1 \times 10^{-2}$ \\
3.7  & $1.2 \times 10^{-2}$ & $1.2 \times 10^{-2}$ \\
3.3  & $6.9 \times 10^{-3}$ & $6.8 \times 10^{-3}$ \\
3.1  & $4.1 \times 10^{-3}$ & $4.2 \times 10^{-3}$ \\
2.7  & $1.1 \times 10^{-3}$ & $8.1 \times 10^{-4}$ \\
2.5  & $4.5 \times 10^{-4}$ & $4.2 \times 10^{-4}$ \\
2.25 & $1.1 \times 10^{-4}$ & $6.7 \times 10^{-5}$ \\
\end{tabular}
\end{table}

\begin{table}
\caption{Summary of characteristic times. The cluster mixing time
$\tau_{\rm cl}$ is computed as in (\ref{ometau}) from the cluster
metric
$\Om_{\rm cl}$. The
single particle time $\tau_D=\sigma^2/(4D)$, the mean time it takes
a particle to diffuse a distance $\sigma$, and the energy mixing
time $\tau_\e$ from (\ref{ometau}) are shown for comparison. (The
values of $\tau_D$ are obtained from the second column in
Table~\ref{table:D}.)}
\label{table:times}
\begin{tabular}{|rrrr|}
$T$ & $\tau_{\rm cl }$ & $\tau_D$ & $\tau_\e$ \\
\tableline
5.5 &   $5 \times 10^2$ &  5.7   & 16 \\
4.6 &   $1 \times 10^3$ &  8.4   & 28 \\
3.7 &   $5 \times 10^3$ &   20   & 69 \\
3.3 &   $2 \times 10^4$ &   36   & 140 \\
3.1 &   $5 \times 10^4$ &   60   & 260 \\
2.7 &   $5 \times 10^5$ &  230   & $1\,100$ \\
2.5 &   $5 \times 10^5$ &  560   & $1\,400$ \\
2.25 &  \hfill $-$      & 2300   & $11\,000$ \\
\end{tabular}
\end{table}

\begin{table}
\caption{Range of fits for the temperature parameter $T_0$ from
least squares fits of the self-diffusion coefficient $D$, the energy
mixing time $\tau_\e$, and the cluster mixing time
$T_{\rm cl}$ to the Vogel-Fulcher
form (\ref{dvf}).}
\label{table:fitD}
\begin{tabular}{|rrr|}
temperature interval & $T_0$ [from $R^2(t)$] & $T_0$ [from
$1/\Om_v(t)$] \\
\tableline
$2.25 \leq T \leq 5.5$ &   1.82 &  1.67 \\
$2.7 \leq T \leq 5.5$  &   1.83 &  1.21 \\
$2.25 \leq T \leq 4.6$ &   0.97 &  1.30 \\
$2.7 \leq T \leq 4.6$  &   0.94 &  1.30 \\
\tableline
                       & $T_\e$ & \\
\tableline
$2.25 \leq T \leq 5.5$ & no fit & \\
$2.7 \leq T \leq 5.5$  & 1.25 & \\
$2.7 \leq T \leq 4.6$  & 1.25 & \\
$3.1 \leq T \leq 5.5$  & 1.68 & \\
$3.1 \leq T \leq 4.6$  & 1.95 & \\
\tableline
 & $T_{\rm cl}$ & \\
\tableline
$2.25 \leq T \leq 5.5$ & no fit & \\
$2.7 \leq T \leq 5.5$  & 1.58 & \\
$2.7 \leq T \leq 4.6$  & 1.57 & \\
$3.1 \leq T \leq 5.5$  & 1.58 & \\
\end{tabular}
\end{table}


\begin{figure}[h]
\caption{Plot of the radial distribution function $g(r)$
versus $r$ at $T=2.7$. Note the split second
peak, a possible signature of a glassy state. Only the
small, majority particles were included. The dimensionless quantity
$r$ is measured in terms of the Lennard-Jones parameter $\sigma$.}
\label{fig:gr}
\end{figure}

\begin{figure}[h]
\caption{Plot of $\Om_v(0)/\Om_v(t)$, the reciprocal
of the velocity fluctuation metric, at  $T=2.7$. Note that
$1/\Om_v(t)$ is not linear for $t < 250\,
\tau$, but becomes approximately linear for longer times. The
temperature $T$ is dimensionless (see text).}
\label{fig:omv}
\end{figure}

\begin{figure}[h]
\caption{Semilog plot of the (dimensionless) self-diffusion
coefficient $D$ versus $1/T$. The data points (filled circles) are
taken from the second column in Table~\ref{table:D}. The solid line
represents the best fit to the results for $D$ in the
interval $2.25 \leq T \leq 5.5$ extracted from
$R^2(t)$ to the Vogel-Fulcher form
$D(T) = A e^{-B/(T-T_0)}$ with $A=0.16$, $B=4.7$, and $T_0 = 1.82$.}
\label{fig:dt}
\end{figure}

\begin{figure}[h]
\caption{Semilog plot of the (dimensionless)
energy mixing time $\tau_\e$ defined in (\ref{ometau}) versus
$1/T$. The data points
(filled circles) are taken from
Table~\ref{table:times}. The solid line represents the best fit to
the Vogel-Fulcher form
$\tau_\e = A\, e^{B/(T-T_\e)}$ with $A=1.24$, $B=9.8$, and
$T_\e = 1.25$ for the six temperatures in the interval $2.7
\leq T \leq 5.5$.}
\label{fig:emetric}
\end{figure}

\begin{figure}[h]
\caption{Plot of the static structure function $S(k)$ at $T=2.7$ as
a function of $k$. The height of the first diffraction maximum
increases and its width decreases as $T$ is lowered from $T=5.5$
(see Table~\ref{table:runs}).}
\label{fig:sk}
\end{figure}

\begin{figure}[h]
\caption{(a) The temperature dependence of $\chi(k_0,T)$,
the height of the diffraction peak of $S(k)$, at $k=k_0 \approx
7.15$. The solid line represents the best fit in the range
$2.7
\leq T \leq 5.5$ and has the form $(T-2.6)^{-0.16}$. (b) The
temperature dependence of $w(k_0,T)$, the width of the diffraction
peak of $S(k)$. The solid line represents the best fit in the range
$2.7 \leq T \leq 5.5$ and has the form
$(T-2.6)^{0.29}$.}
\label{fig:chi}
\end{figure}

\begin{figure}[h]
\caption{Plot of $P(h)$, the estimated probability density
of $h$, the relative variance of the edge length of the Voronoi
polygons, at $T=5.5$ (solid line) and $T=2.7$ (dotted line). The
data points are not shown to avoid confusion. Regular hexagons and
hence lower values of $h$ are much more likely at low temperatures.
Note that $P(h)$ is not bimodal at any temperature. The shoulder in
$P(h)$ at $h\approx 0.25$ appears to be real rather than
statistical error.}
\label{fig:ph}
\end{figure}

\begin{figure}[h]
\caption{A typical configuration at $T=3.3$ showing the Voronoi
polygon for each particle and the clusters of the majority small
particles (shaded hexagons).}
\label{fig:config}
\end{figure}

\begin{figure}[h]
\caption{Log-log plot of $n_s$, the cluster size distribution,
versus size $s$ at $T=3.3$ for $N=500$ particles. The solid
line with slope $x = 1.85$ is the best fit in the range $6 \leq s
\leq 61$.}
\label{fig:ns3.3}
\end{figure}

\begin{figure}[h]
\caption{Log-log plot of $n_s$
versus $s$ at $T=3.1$ for $N=500$ particles. The results for the
first $50\,000\,\tau$ (open circles) and the second $50\,000\,\tau$
(open squares) are shown separately. The system was run for
$50\,000\, \tau$ before data was taken. Note the
difference in the size distribution of clusters for larger values
of $s$.}
\label{fig:ns3.1}
\end{figure}

\begin{figure}[h]
\caption{Log-log plot of $n_s$ versus $s$ at $T=5.5$ for $N=500$
particles. The plot shows curvature indicating that there is not a
simple scaling regime. The solid line is a fit to the form
$n_s = A s^{-x}e^{-s/m_s}$ with $A=0.038$, $x = 1.1$, and $m_s=13$.
}
\label{fig:ns3.7}
\end{figure}

\begin{figure}[h]
\caption{Log-log plot of $n_s$
versus $s$ at $T=2.7$ for $N=20\,000$ particles. The linear fit was
done for $5 \leq s \leq 61$ and yields a slope of $x=1.75$.}
\label{fig:ns20000}
\end{figure}

\begin{figure}[h]
\caption{The time-dependence of $\Omega_{\rm cl}(0)/\Omega_{\rm
cl}(t)$, where $\Omega_{\rm cl}(t)$ is the cluster fluctuation
metric defined in (\ref{omcl}) at (a) $T = 5.5$ and (b) $T = 3.1$.
Note the very different vertical scales in (a) and (b). The time
$t$ is given in terms of $\tau$ defined in the text.}
\label{fig:omcl}
\end{figure}

\begin{figure}[h]
\caption{Plot of the cluster mixing time $\tau_{\rm cl}$ (solid
circles) versus
$T$, where $\tau_{\rm cl}$ is extracted from the linear
time-dependence of
$1/\Omega_{\rm cl}$. The
solid line represents the best fit to the Vogel-Fulcher form
$\tau_{\rm cl}(T) = A\, e^{C/(T-T_{\rm cl})}$ with $A=19.8$,
$C=11.8$, and
$T_{\rm cl} = 1.58$ in the
interval $3.1
\leq T
\leq 5.5$.}
\label{fig:tau_cl}
\end{figure}

\begin{figure}[h]
\caption{The temperature-dependence of the (a) (dimensionless)
enthalpy $H$ and (b) the heat capacity $C_P$ at $P=70$. Note that
$C_P$ has a small peak at
$T \approx 4$. As explained in the text, the slope of $H(T)$ was
computed at each value of
$T=T_i$ and the value of $C_P$ at $T=T_i$ was computed by doing a
least squares fit to 15 successive slopes in the interval $T_{i+7}
- T_{i-7}$.}
\label{fig:appen}
\end{figure}


\begin{thebibliography}{1}

\bibitem{reviews} Reviews of the properties of structural
glasses can be found for example in J. P. Hansen, D. Levesque, and
J. Zinn-Justin, eds., {\sl Liquids, Freezing, and the Glass
Transition}, North-Holland (1991); W. G\"otze and L. Sj\"ogren,
{\sl Rep. Prog. Phys.} {\bf 55}, 241 (1992); R. D. Mountain, {\sl
Int. J. Mod. Phys.} C {\bf 5}, 247 (1994); C. A. Angell, {\sl
Science} {\bf 267}, 1924 (1995); S. C. Glotzer, ed., {\sl Glasses
and the Glass Transition} in {\sl Comput. Mat. Sci.} {\bf 4}, 283
(1995); M. D. Ediger, C. A. Angell, and S. R. Nagel, {\sl J. Phys.
Chem.} {\bf 100}, 13200 (1996).

\bibitem{prl} A. I. Mel'cuk, R. A. Ramos, H. Gould, W. Klein,
and R. D. Mountain, {\sl Phys. Rev. Lett.} {\bf 75}, 2522 (1995).

\bibitem{clumps} W. Klein, H. Gould, R. Ramos, I. Clejan, and A.
Mel'cuk, {\sl Physica} A {\bf 205}, 738 (1994).

\bibitem{kac} M. Kac, G. E. Uhlenbeck, and P. C. Hemmer, {\sl J.
Math. Phys.} {\bf 4}, 216 (1961).

\bibitem{grewe} N. Grewe and W. Klein, {\sl J. Math. Phys.} {\bf
18}, 1729, 1735 (1977).

\bibitem{ramos} R. A. Ramos, Ph.D. thesis, Boston University
(1994).

\bibitem{dieter} D. W. Heermann, W. Klein, and D. Stauffer, {\sl
Phys. Rev. Lett.} {\bf 50}, 1062 (1983).

\bibitem{novotny} M. A. Novotny, W. Klein, and P. Rikvold, {\sl Phys.
Rev.} B {\bf 33}, 7729 (1986) and M. A. Novotny, P. Rikvold and W.
Klein, unpublished.

\bibitem{clejan} I. Clejan, Ph.D. thesis, Boston University (1994).

\bibitem{thirumalai} D. Thirumalai and R. D. Mountain, {\sl Phys.
Rev.} E {\bf 47}, 479 (1993); R. D. Mountain and D. Thirumalai, {\sl
Physica} A {\bf 210}, 453 (1994). In this context a system is
ergodic if time averages and ensemble averages are equal.

\bibitem{wong} Y. J. Wong and G. W. Chester, {\sl Phys. Rev.} B {\bf
35}, 3506 (1987).

\bibitem{gt} See for example, H. Gould and J. Tobochnik, {\sl
Introduction to Computer Simulation Methods}, second edition,
Addison-Wesley (1996) or D. Rapaport, {\sl The Art of Molecular
Dynamics Simulation,} Cambridge University Press (1996).

\bibitem{monette} L. Monette and W. Klein, private
communication; D. Stauffer, private communication.

\bibitem{coniglio} A. Coniglio and W. Klein,
{\sl J. Phys.} A {\bf 13}, 2775 (1980).

\bibitem{klein} W. Klein, {\sl Phys. Rev. Lett.} {\bf 65}, 1462
(1990).

\bibitem{clark} M. A. Glaser and N. A. Clark, {\sl Adv.
Chem. Phys.} {\bf 83}, 543 (1993), use sixfold bond orientational
order to define the clusters in a one component,
two-dimensional Lennard-Jones system.

\bibitem{voronoi} R. E. M. Moore and I. O. Angell, {\sl J. Comp.
Phys.} {\bf 105}, 301 (1993).

\bibitem{melcuk} A. I. Mel'cuk, Ph.D. thesis, Clark University
(1994).

\bibitem{medvedev} N. N. Medvedev, A. Geiger, and W. Brostow, {\sl
J. Chem. Phys.} {\bf 93}, 8337 (1990).

\bibitem{mountain2} R. D. Mountain in {\sl Symposium Series No. 676,
Supercooled Liquids: Advances and Novel Applications}, J. Fourkas,
D. Kivelson, K. Nelson, and U. Mohanty, eds. (American Chemical
Society, Washington, D. C., 1997), Chapter 9; W. Kob, C. Donati, S.
J. Plimpton, P. H. Poole, and S. C. Glotzer, {\sl Phys. Rev.
Lett.} {\bf 79}, 2827 (1997).

\bibitem{ediger} See for example, M. T. Cicerone and M. D. Ediger,
{\sl J. Chem. Phys.} {\bf 103}, 5684 (1995).

\bibitem{kleinpc} W. Klein et al., private communication.

\bibitem{lovett} R. Lovett, {\sl J. Chem. Phys.} {\bf 66}, 1225
(1977).

\bibitem{unger} W. Klein and C. Unger, {\sl Phys. Rev.} B {\bf 28},
445 (1983); C. Unger and W. Klein, {\sl Phys. Rev.} B {\bf 29},
2698 (1984).

\bibitem{leyvraz} W. Klein, and F. Leyvraz, {\sl Phys. Rev. Lett.}
{\bf 57}, 2845 (1986).

\bibitem{yang} J. Yang, H. Gould, W. Klein, and R. Mountain, {\sl J.
Chem. Phys.} {\bf 93}, 711 (1990).

\bibitem{desai} R. C. Desai and A. R. Denton, {\sl Growth and
Form}, H. E. Stanley and N. Ostrowsky, eds., Martinus Nijhoff Press
(1986).

\bibitem{lou} L. Colonna-Romano, Ph.D. thesis, Clark University, in
preparation.

\bibitem{dixon} P. K. Dixon, L. Wu, S. R. Nagel, B. D. Williams, and
J. P. Carini, {\sl Phys. Rev. Lett.} {\bf 65}, 1108 (1990).

\bibitem{menon} N. Menon and S. R. Nagel, {\sl Phys. Rev. Lett.} {\bf
74}, 1230 (1995).

\bibitem{binder} K. Binder, {\sl Phys. Rev.} A {\bf 29},
341 (1984).

\bibitem{kivelson} S. A. Kivelson, A. Zhao, D. Kivelson, T. M.
Fischer, and C. M. Knobler, {\sl J. Chem. Phys.} {\bf 101}, 2391
(1994); D. Kivelson, S. A. Kivelson, X. Zhao, Z. Nussinov, and G.
Tarjus, {\sl Physica} A {\bf 219}, 27 (1995).

\bibitem{mountain} R. D. Mountain, {\sl J. Chem. Phys.} {\bf 102},
5408 (1995). 

\bibitem{grest} R. M. Ernst, S. R. Nagel, and G. S. Grest, {\sl Phys.
Rev.} B {\bf 43}, 8070 (1991).

\bibitem{wahnstrom} G. Wahnstr\"om, {\sl Phys. Rev.} A {\bf 44},
3752 (1991).

\bibitem{andersen} H. C. Andersen, {\sl J. Chem. Phys.} {\bf 72},
2384 (1980).

\end{thebibliography}
\end{document}